\documentclass[iop]{emulateapj}

\shorttitle{Hard X-ray footpoint sizes and positions}
\shortauthors{Battaglia \& Kontar}
\begin{document}

\title{Hard X-ray footpoint sizes and positions as diagnostics of flare accelerated energetic electrons
in the low solar atmosphere}

\author{M. Battaglia and E. P. Kontar}
\affil{SUPA, School of Astronomy and Astrophysics, University of Glasgow, G12 8QQ, UK}
\email{e-mail: Marina.Battaglia@glasgow.ac.uk}

\begin{abstract}
The hard X-ray (HXR) emission in solar flares comes almost exclusively from a very small
part of the flaring region, the footpoints of magnetic loops.
Using RHESSI observations of solar flare footpoints,
we determine the radial positions and sizes of footpoints as a function of energy
in six near-limb events to investigate the transport of flare accelerated electrons
and the properties of the chromosphere. HXR visibility forward fitting allows
to find the positions/heights and the sizes of HXR footpoints along
and perpendicular to the magnetic field of the flaring loop at different
energies in the HXR range. We show that in half of the analyzed events,
a clear trend of decreasing height of the sources with energy is found.
Assuming collisional thick-target transport, HXR sources
are located between $600$ and $1200$~km above the photosphere for
photon energies between $120$ and $25$ keV respectively.
In the other events, the position as a function of energy is constant within the
uncertainties. The vertical sizes (along the path of electron propagation) range from 1.3 to 8 arcseconds which is up to a factor 4 larger than predicted by the thick-target
model even in events where the positions/heights of HXR sources
are consistent with the collisional thick-target model.
Magnetic mirroring, collisional pitch angle scattering and X-ray albedo
are discussed as potential explanations of the findings.
\end{abstract}


\section{Introduction}
In the traditional flare model particles are accelerated in the corona then precipitate
along the field lines of a magnetic loop to the chromosphere where they are stopped
producing bremsstrahlung emission in the process. In the classical thick-target
model \citep{Br71}, it is assumed that collisional interaction of fast electrons
with the ambient plasma leads to energy-loss, while other mechanisms such as pitch
angle scattering or mirroring of the electrons in a converging magnetic
field are neglected. It is therefore expected that electrons with higher energies
penetrate deeper into the chromosphere before they are fully stopped.
The stopping depth depends on the initial electron energy and the ambient density. Expressed in terms of the column depth $N(s)=\int n(s)\mathrm{d}s$, where $n(s)$
is the ambient density along the electron path, the stopping depth
is given as $N_{stop}= E_0^2/2K$, where $E_0$ is the initial
energy of the accelerated electron and $K=2\pi e^4 \Lambda$ \citep{Br72, Br02}. The Coulomb logarithm $\Lambda$
has typical values of $\sim 20$ in the (ionized) corona and $\sim 7$ in the (neutral)
chromosphere \citep{Br73,Em78}. For an electron flux distribution $F(E,s)$
with energy $E$ at distance $s$ from the point of injection,
the observed X-ray flux at Earth is:
\begin{equation} \label{beq}
I(\epsilon,s)=\frac{n(s)A(s)}{4\pi R^2}\int_{\epsilon}^{\infty}F(E,s)\sigma(\epsilon,E)\mathrm{dE},
\end{equation} where $n(s)$ is the density, $A(s)$ the width of the magnetic flux tube
at distance $s$, $R$ the Sun-Earth distance and $\sigma(\epsilon,E)$ the angle-averaged
bremsstrahlung cross-section \citep{Ha97}. The isotropic approximation of emission is supported by statistical observations \citep{Ka88,Ves87, Ka07} and more recent HXR observations using albedo in imaging \citep{Ba11} and spectroscopy \citep{Ko06a}. For increasing density along $s$, Eq.~\ref{beq}
has a maximum for a given photon energy $\epsilon$, i.e. the observed HXR emission
will have a maximum at a certain chromospheric depth, depending on energy.

Observational evidence for height dependent HXR sources was found early on in stereoscopic
observations \citep{Ka83} and in a statistical way using Yohkoh \citep{Mat92}.
\citet{Ka83} derived that HXR sources for energies $>150$~keV should be
at heights less than $2500$~km above the photosphere.
\citet{Fl96} used test particle simulations to find the expected position
as a function of energy including collisional pitch angle scattering
and magnetic mirroring and compared the results with observations from Yohkoh.
This study demonstrated the influence of a converging magnetic field
on the height of the X-ray source, showing that the Yohkoh
observations of $53-93$ keV sources at average
heights $\sim 6000$ km presented by \citet{Mat92} are consistent
with partial trapping of electrons in a magnetic loop. However,
the use of H$\alpha$ flare locations as the reference for height estimates
could be the reason for the large heights found by \citet{Mat92}.

The high spatial resolution of RHESSI \citep{Li02,Hur02} now makes it possible to study
individual events with higher accuracy \citep{As02, Mr06, Liu06}.
More recently \citet{Ko08} and \citet{Koet10} analyzed a limb event using the newly
developed visibility technique \citep{Sc07}. This allows for measurements of HXR source
positions with sub-arcsecond resolution. \citet{Pr09} and \citet{Pe10}
went one step further and computed the electron distributions
found from inversion of the X-ray visibilities.
In all those studies a decrease of the radial position of the sources
with increasing energy is found. \citet{Br02} showed how this can be used
to determine the chromospheric density structure. This was applied by \citet{As02}
to an event on 2002 February 20 and to a limb event on 2004 January 6 by \citet{Koet10}
who found their observations to be consistent with an exponential
chromospheric density profile with scale height $\approx$ 150 km.
The HXR source heights have been found to decrease
from $\approx 1200$ km at 20 keV to $\approx 700$ at 160 keV.
A statistical survey of over 800 flares \citep{Sa10} found similar heights
and that HXR sources appear within a relatively narrow range of heights of $0.5$~Mm.

Due to constantly improving analysis techniques for RHESSI data,
it is now not only possible to determine the position of footpoint sources
with high accuracy, but also the characteristic sizes at different energies
and hence at different heights \citep{Ko08}. Moreover, it is possible to assess
the extent of footpoint sources parallel and perpendicular
to the magnetic loop, as demonstrated by \citet{Koet10}.
This opens up a completely new approach to the diagnostics of the magnetic field
structure of the chromosphere and the physics of electron transport
in the chromosphere.
In a thick-target case, electrons with higher energies penetrate deeper
into the chromosphere. In a converging magnetic loop, the higher energetic
electrons will therefore penetrate into regions with higher
magnetic field strength. This should be reflected in the horizontal
size of the X-ray source which is expected to decrease
with energy. Such a behavior was indeed observed in the event
analyzed by \citet{Ko08}, who also inferred a magnetic scale height
of the order of 300 km. At the same time the vertical extent
of a source should be determined by the same physics
of electron transport. However, \citet{Koet10} find that the vertical
sizes of HXR sources are inconsistent with the sizes expected
from the collisional thick-target. While they interpret this in terms
of a multi-threaded chromosphere with different density profiles along
the different threads, alternative explanations within a monolithic chromosphere
framework such as collisional pitch angle scattering or magnetic mirroring
are possible.  Another effect that influences the observed positions
and sizes of X-ray sources is X-ray albedo, as shown by \citet{Ko10}.
\citet{Ba11} went a step further and demonstrated how the measured
full width half maximum (FWHM) in simulated and observed maps can be used
to constrain the true source size and the directivity.

In this paper, we investigate the energy dependence of not only the position,
but the sizes of footpoints in a carefully selected sample of six limb events
observed by RHESSI. We used visibility forward fitting to find the positions
and FWHM sizes of footpoints as a function of energy. We show that in half
of the observed sources, the position decreases with energy and can be interpreted in a simple thick-target approach. In the other events the position is constant
within the uncertainties. Contrary to the predictions from the
collisional thick-target model, the vertical sizes along the electron
path are a factor of 2-4 larger and weakly decrease with energy
for all flares analyzed. Possible explanations
include collisional pitch angle scattering, magnetic
mirroring and X-ray albedo.

\begin{figure*}[!]
\begin{center}
\resizebox{0.85\hsize}{!}{\includegraphics{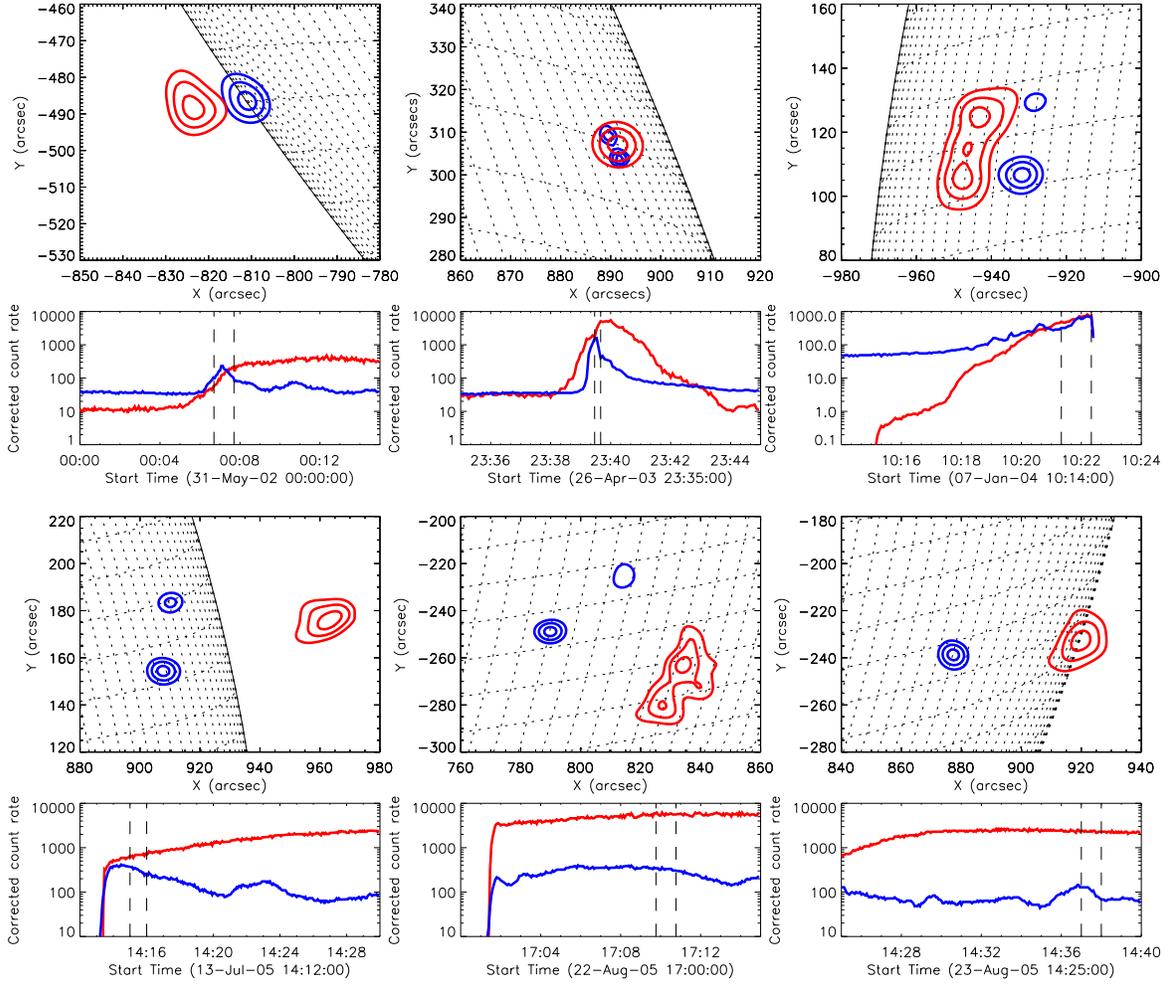}}
\end{center}
\caption {Images and lightcurves of the selected events. Contours of 50\%, 70\% and 90\% of the maximum emission in CLEAN images using detectors 3-8 (1-8 for the high energies in the 26 April event) are given for energy ranges 6-12 keV (red, thermal emission) and 35-55 keV (blue, non-thermal emission). Lightcurves of corrected count rates in the 6-12 keV (red) and 25-50 keV (blue) energy bands are displayed underneath each image. The dashed lines mark the analyzed time interval. }
\label{plotone}
\end{figure*}

\section{Flare selection}
The primary selection criterion were events with unambiguous footpoint sources
observed to high energies. We restricted our selection to events of GOES class M
and above in order to have high enough count rates. At the same time
events with strong pulse pile-up (live-time less than $\approx$ 85\%) were excluded \citep{Smith02}.
Pile-up appears when two or more low energy photons,
mostly from the coronal source, are detected as a single
photon with higher energy. It is of less concern in footpoint analysis, as the pile-up signature
in images appears as artificial high energy source
in the corona where the bulk of the soft X-ray emission originates. Except for the 2005 August 22 the attenuator state during the events presented here was 1. Thus the peak of the pile-up is at 24 keV, below our analyzed energies. In the 2005 August 22 (attenuator state 3), the live-time was high enough for pile-up to be minimal. To test the effect potential pile-up counts would have on the visibility forward fit, we fitted the two footpoints in the 2005 July 13 event in an energy-band of 16-40 keV in which emission from the coronal source is present, mimicking pile-up. It is found that the positions are shifted by less than 0.5 arcsecond and the FWHM change by less than 0.2 arcsec, both effects being smaller than the uncertainties in the fit parameters even though the intensity of the coronal emission per unit area was only a factor of 2 less than the footpoint emission per area.
Further, to reduce projection
effects and find events with a clear morphology,
the event selection was restricted to limb events (angular offset from Sun center larger
than 700 arcsec). This also minimizes the influence of X-ray
albedo on the source positions and sizes (Section \ref{physics}).
Quick look images provided by the HESSI Experimental Data Center
\citep[HEDC\footnote{Unavailable now as it went off-line at the beginning of 2010},][]{HEDC}
were used to search for events with one or two footpoints visible up to at least
about 80 keV. Events with more than two footpoints or other complex structures
such as the 2002 July 23 event \citep{Em03} were excluded.
Finally, six events satisfying aforementioned criteria were
selected (The key parameters of the flares are listed in Table~\ref{events}).
\begin{deluxetable*}{lllcl}
\tabletypesize{\scriptsize}
\tablecaption{Dates, times, GOES class, electron spectral index $\mathbf{\delta}$ and cosine of heliocentric angle $\theta$ of the flaring region for the analyzed flares. The times given are the start times of the 1-minute observation interval (Section \ref{datanalysis}), except for the 2003 April 26 event
for which the full interval (less than 1 minute) is given.}
\tablehead{
\colhead{Date} & \colhead{Start time [UT]} & \colhead{GOES class} & \colhead{spectral index $\mathbf{\delta}$} &\colhead{$\mathbf{\cos \theta}$} }
\startdata
2002 May 31 &00:06:42&M2.5& 3.5 & 0.04\\
2003 Apr 26&23:39:28-23:39:40&M1.1&3.6&0.16\\
2004 Jan 7&10:21:20&M8.3 &3.4&0.28\\
2005 Jul 13&14:15:00&M5.1&4.1&0.2\\
2005 Aug 22&17:09:46&M6.3 &3.7&0.47\\
2005 Aug 23&14:37:00&M3.0 &4.5&0.29\\
\enddata
\label{events}
\end{deluxetable*}

Figure~\ref{plotone} gives the temporal and morphological
overview of the events. Contours at 50 \%, 70\% and 90 \% of the maximum emission
in CLEAN images are shown for energy bands 6-12 keV and 35-55 keV.
The 6-12 keV emission represents the thermal (coronal) emission,
while the 35-55 keV emission indicates the footpoint emission.
Two of the events only have one footpoint.
Two events have one strong and one weak footpoint while in the other two events,
the intensities of HXR footpoint emission are comparable.
The lightcurves of corrected count rates in the 6-12 keV and 25-50 keV energy bands are presented underneath the images.

\section{Data analysis} \label{datanalysis}
For the analysis we chose a time interval of 1 minute during which the footpoint emission
is observed to highest energies in the quicklook images on HEDC (except in the 2002 April 26 event during
which attenuator state changes only allowed an interval as long as 12 sec). This does not necessarily coincide with the peak of the HXR emission.
Shorter time intervals generally result in poorer signal-to-noise
ratio for HXR visibilities. Further, the longer time interval improves
the aspect phase coverage because of the precession
of the RHESSI spin axis (see \ref{visfwdfit})
and also helps to minimize the effect of data gaps \citep{Hur02}.
On the other hand it is often observed that footpoints move spatially
over the course of one minute \citep[e.g.][]{Gr05a,Fl04,Kr03}, which might affect
the observed positions and sizes. In the events presented here,
the motion of the sources is found to be less than one arcsec
during the observed time interval. This is of the order of uncertainties
in the measurements of the positions and sizes. Therefore,
this effect can be neglected.

Images at several energy bands in the non-thermal energy range were made
to find the positions and the sizes of the footpoints.
The energy-binning was chosen to provide 5 - 6 images in the hard X-ray domain per event,
starting from 25 keV (Detector 2 sensitivity threshold, see \citealt{Smith02}) up to the highest observed
energies (100 - 200 keV, typically). A finer energy binning would be desirable
but would lead to insufficient number of counts in the individual energy bands.
We used CLEAN \citep{Hog74,Hur02} and Pixon \citep{Pi92,Me96} for a first impression
of the sources, their intensity and the shape. The positions and sizes
as a function of energy were then found using the technique
of visibility forward fitting \citep[e.g.][]{Ba11,De09,Xu08,Ha08}.

\subsection{Visibility forward fitting}\label{visfwdfit}
\begin{figure*}[!]
\center{\resizebox{0.85\hsize}{!}{\includegraphics{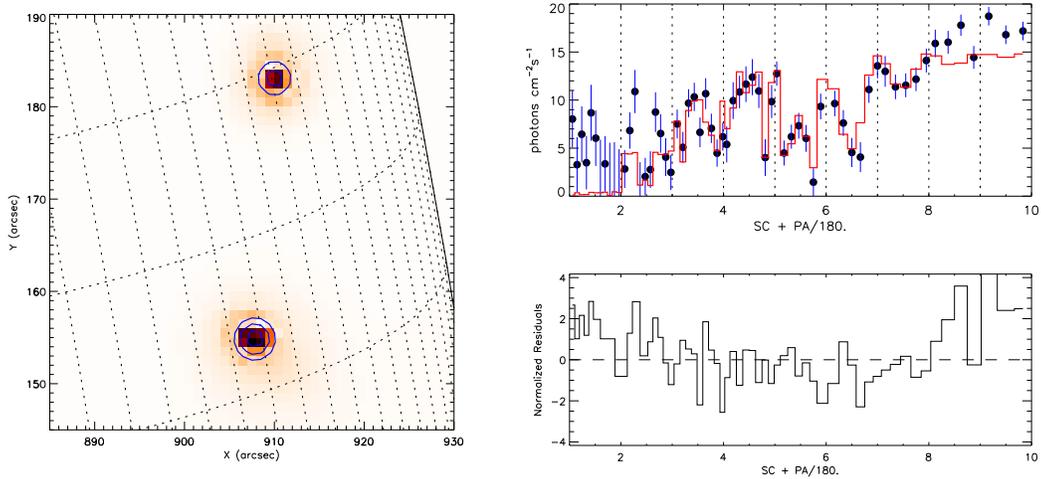}}}
\caption {Illustration of visibility forward fit. Left: Pixon map overlaid with
the 50\%, 70\% and 90\% contours of the source model (two circular Gaussians).
Top right: Measured visibility amplitudes (black dots) with statistical
uncertainties (blue) as a function of position angle for each RHESSI
detector (1 to 9). Bottom right: Normalized residuals
of the visibility amplitudes. }
\label{plottwo}
\end{figure*}

Visibilities are a concept widely used in radio astronomy. In recent years it has been
adapted to be used with RHESSI \citep{Hur02,Sc07}.
HXR visibility forward fitting is ideally suited to find positions
and sizes of flare sources for several reasons.
X-ray visibilities $V(u,v)$ are the 2D spatial Fourier components
of the X-ray distribution $I(x,y)$:
\begin{equation} \label{vis}
V(u,v)=\int_x\int_yI(x,y)\exp[{2\pi i(ux+vy)]}\mathrm{d}x\mathrm{d}y,
\end{equation}
so the total flux, position and size of an X-ray source (comp. Eq.~\ref{beq})
can be straightforwardly expressed through the 0th, 1st and 2nd moment
of the X-ray distribution. Measurements of the 1st and 2nd moment
from reconstructed images (e.g. CLEAN, Pixon) are impractical
since the result is dependent on the selection of an image region.
Because RHESSI images suffer from reconstruction
noise and are hampered by imaging artefacts
such as side lobes from the CLEAN beam, so will the moments.
Visibility forward fitting does not require the reconstruction
of an image itself and is therefore the most direct way to
find a measure of the moments \citep{Ba11}. If the observed source has a Gaussian shape then the forward fit parameters will represent the true moments. Further, statistical errors
for the fit parameters are computed from the visibility errors
making visibility forward fitting the only method that
provides error estimates for the measured parameters.
Figure \ref{plottwo} gives an example of a visibility
forward fit. All detectors were used since we only analyze energies above the sensitivity threshold of detector 2 ($>$ 25 keV). On the left-hand side a Pixon map is shown,
overlaid with the 50\%, 70\% and 90 \% contours
of the source model (two circular Gaussians).
The top panel on the right-hand side shows the observed
visibility amplitudes with statistical uncertainties and the fitted model. The bottom panel on the right-hand side
displays the normalized residuals of the visibility amplitudes. Note that the model is fitted
using all $V(u,v)$, not only the amplitudes $|V(u,v)|$ shown in Figure \ref{plottwo}.
Apart from the source shape given by the model, the number of roll-bins i.e. the spatial
coverage for which visibilities are computed, has to be chosen.
A large number of roll bins will provide finer spatial coverage.
If used, especially for the finer grids (detectors 1-3), potential ellipticity
of the source can be better assessed and fitted, so the visibilities might
even reflect smaller source structures. At the same time, the errors of the individual
visibilities tend to be larger for small count rates as the same number of counts
is spread over a larger number of visibilities. A smaller number of roll bins provides
coarser spatial coverage but with smaller errors of the individual visibilities.
However, the choice of roll bin number does not affect the inferred forward fit parameters (flux, position and size), as tests using different settings proved.
\subsection{Spectroscopy}
\begin{figure*}[!]
\center{\resizebox{0.85\hsize}{!}{\includegraphics{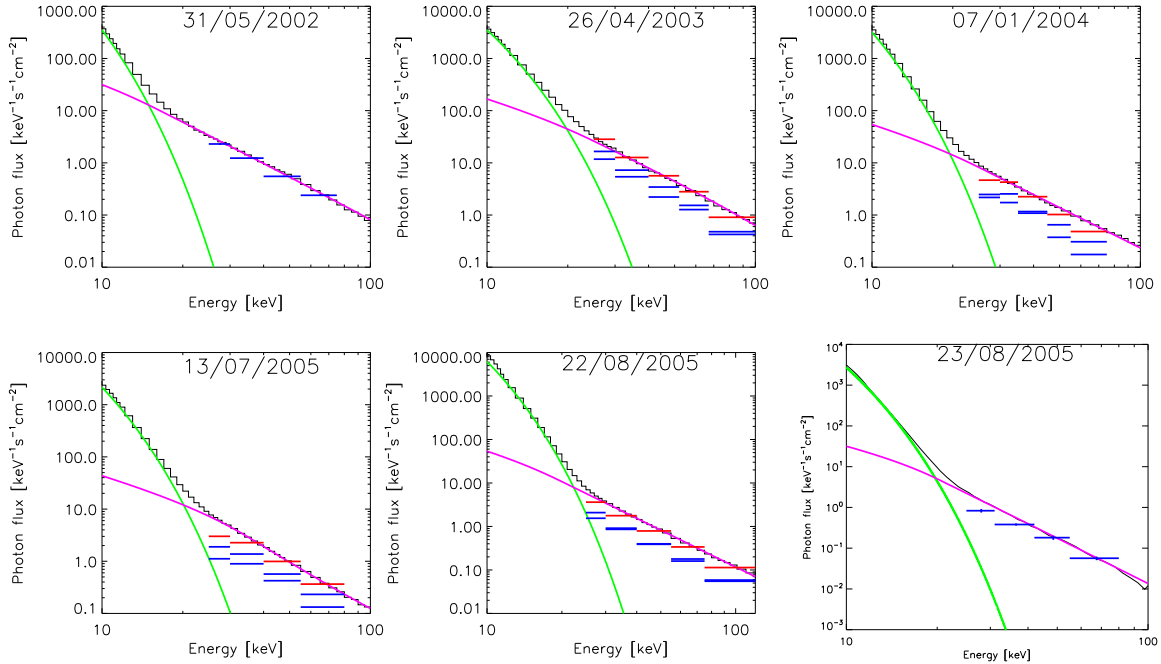}}}
\caption {Full Sun spectra, fitted with a thermal model (green) and a non-thermal power-law (purple). The blue lines indicate the flux from the individual sources found in visibility forward fitting. The red lines give the total of the individual sources in the case of two footpoints. }
\label{plotthree}
\end{figure*}
Full Sun spectra were analyzed for all events to gather information about the thermal and non-thermal contribution to the emission.
This allows to determine the lower limit of non-thermal energies that can be analyzed
in imaging without the risk of contamination from the coronal source.
The spectra of all events were fitted with a thermal model
plus a thick-target power-law model (Fig. \ref{plotthree}).
This provides the electron spectral index $\delta$ of accelerated or into the thick-target injected 
electrons that is used as a parameter in fitting the chromospheric density.
The comparison of the flux from full-Sun spectra with the total flux from the footpoints
as found in visibility forward fitting is an independent test
of the consistency of the analysis (Fig. \ref{plotthree}).
Assuming the HXR emission comes predominately from the footpoints,
the total flux from the individual sources equals the flux from the full-Sun spectra
if it has been accounted for correctly. The full-Sun spectra are shown
in Fig. \ref{plotthree} with the flux from visibility forward fitting overlaid.
Generally there is a good agreement
between the full Sun flux and the total footpoint flux found in visibility forward fitting.
The exceptions are the points at the lowest energies in the events of 2005 July 13
and 2005 August 23, which could be related to some non-footpoint emission
from the coronal source, albedo or the loop itself.
\section{Results}
\begin{figure*}[!]
\center{\resizebox{0.85\hsize}{!}{\includegraphics{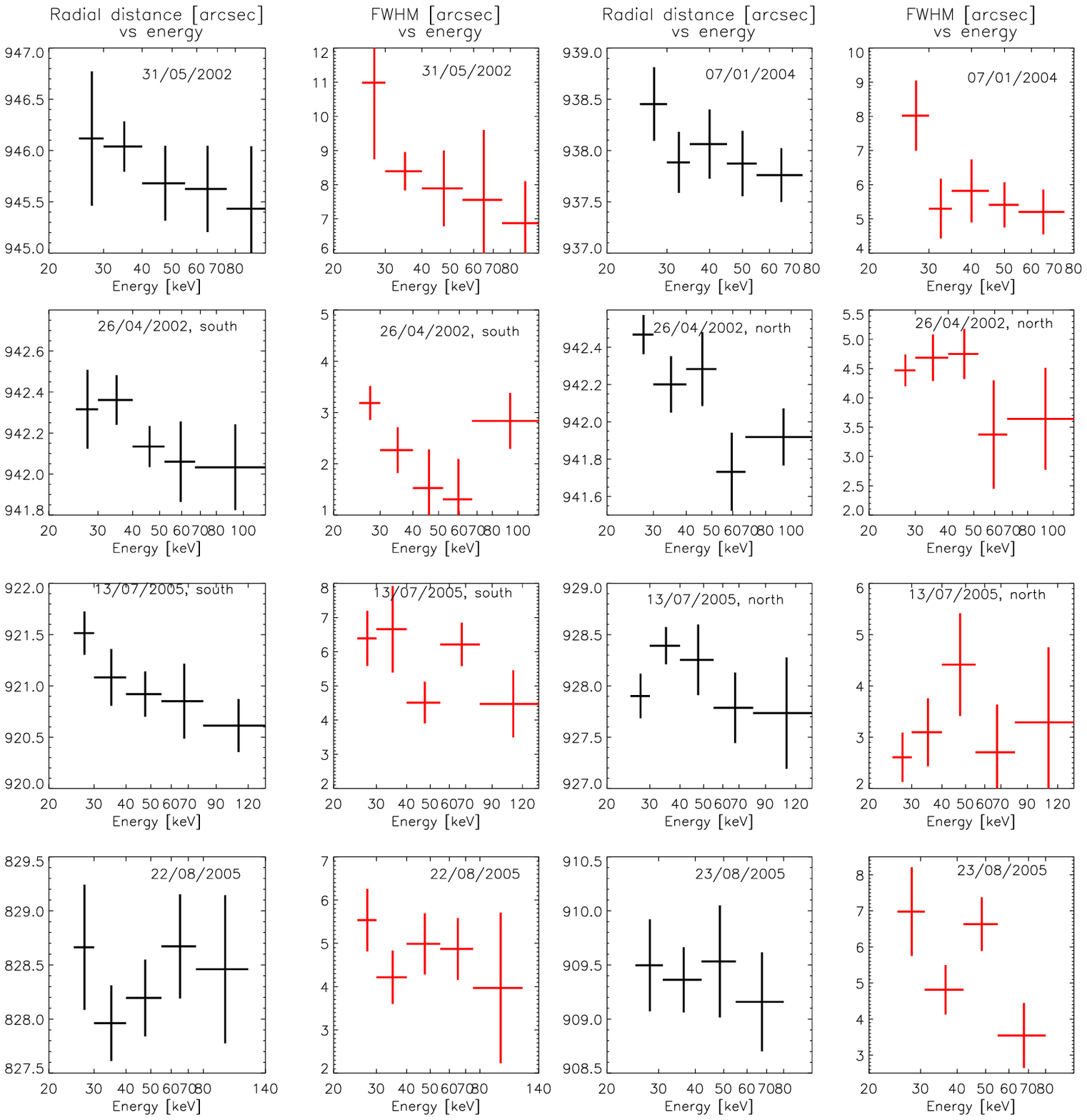}}}
\caption {Results from visibility forward fitting for all events.
First and third column: Radial distance as a function of energy.
Second and fourth column: FWHM as a function of energy.}
\label{plotfour}
\end{figure*}
Although all flares have a number of common features, some aspects are highly individual
and it is difficult to make general statements that hold for all. We therefore make some general
remarks on the results and explain how the chromospheric density was inferred
from the measurement of the position as a function of energy.
Then, each event is discussed individually in more detail.
Figure \ref{plotfour} presents the measurements of radial position and source size for all events.
The first and third column illustrate the measured radial distance (2D distance measured from the solar disk centre) as a function of energy. The second and fourth column show the size as a function of energy.

\subsection{Radial position and density}\label{subsdensity}
Visibility forward fitting returns the $x-y$ position (1st moment) of the fitted source including
errors as a function of energy ($x(\epsilon)$,$y(\epsilon)$). The radial position can readily
be found as $r(\epsilon)=\sqrt{x(\epsilon)^2+y(\epsilon)^2}$. In the classical thick-target
model HXR emission at higher energies is expected to originate deeper in the chromosphere.
This can be used to find the chromospheric density profile by fitting a model to the observed
positions. We use the same approach and general assumptions as \citet{Koet10}.
A hydrostatic density profile of the form
\begin{equation} \label{denseq}
n(h)=n_c+n_0\exp\left(-\frac{h}{h_0}\right)
\end{equation}
is assumed where $h_0$ is the density scale height, $n_0$ the photospheric density and $\mathbf{n_c=9\times 10^9}$ $cm^{-3}$ is the coronal density, taken as constant. The density profile is illustrated in Fig. \ref{plotfive}.
Based on \citet{Ver81} we chose the fixed value of $n_0=1.16\times 10^{17}$ cm$^{-3}$
for $h=0$ (photospheric level). Assuming a vertical loop, the height of the source
above the photosphere, $h(\epsilon)$ can be expressed through the radial distance
from the Sun center as $h(\epsilon)=r(\epsilon)-r_0$ where $r_0$ is the radial distance that corresponds to the photospheric height $h=0$.
The parameters $h_0$ and $r_0$ can be found by forward fitting the measured
radial distances $r(\epsilon)$ with the density model in Eq.~\ref{denseq}.

\subsection{Size and shape}
\begin{figure}[h]
\center{\resizebox{0.85\hsize}{!}{\includegraphics{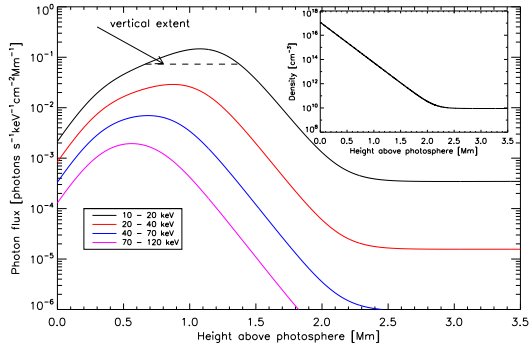}}}
\caption {Illustration of the photon flux as a function of height above the chromosphere for different energy bands (explained in the legend) in a simple thick-target model. The width of the curves at the level of half the peak flux gives the vertical size of the source. Inset: Density profile.} 
\label{plotfive}
\end{figure}

\begin{figure}
\center{\resizebox{0.85\hsize}{!}{\includegraphics{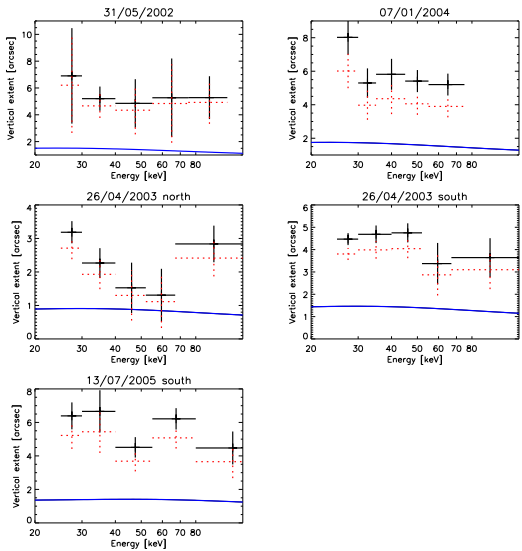}}}
\caption {Vertical extent (FWHM in case of circular Gaussian sources, FWHM of minor axis
in the case of the 2002 May 31) as a function of energy. The red points indicate the projection corrected extents (Sect. \ref{pro}). The blue line
gives the expected vertical extent from the density model that was found
by fitting the radial positions of the respective events.  }
\label{plotsix}
\end{figure}

The second parameter of interest is the source size. In visibility forward fitting one or two Gaussian sources 
\begin{equation}
I(x,y)\sim \exp{\left[-\left(\frac{x^2}{2\sigma_a^2}+\frac{y^2}{2\sigma_b^2}\right)\right]}
\end{equation}
are fitted, where $\sigma_a$ and $\sigma_b$ are the standard deviations. Visibility forward fit returns the FWHM and the eccentricity $e$ from which the major and minor axes ($a=2\sqrt{2\ln 2}\sigma_{a}$ and $b=2\sqrt{2\ln 2}\sigma_{b}$) as a function of energy can be found: $a(\epsilon)=FWHM(\epsilon)(1-e(\epsilon)^2)^{-1/4}$ and $b(\epsilon)=FWHM(\epsilon)(1-e(\epsilon)^2)^{1/4}$. In the case of a circular Gaussian fit, the FWHM is simply the diameter at half the peak-flux level and $a=b$. If the fitted source is a true Gaussian in reality, the size is directly related to the second moment of the X-ray flux distribution \citep{Ba11}.
%
Depending on the loop geometry the major and minor axes are the sizes perpendicular and parallel to the magnetic field of the loop \citep[as in the event analyzed by][]{Koet10}. In this case, the minor axis is measured radially from Sun center and represents the size vertical to the photosphere while the major axis is measured perpendicular to the radial direction, i.e. parallel to the photosphere.
In a simple thick-target model, the vertical size depends on the density structure
and the electron spectral index. Using the density model found as described
in Section~\ref{subsdensity}, the expected X-ray source profile along the loop
can be calculated for different energies. This is illustrated in Fig.~\ref{plotfive}.
The expected photon flux as a function of height above the photosphere
is shown for different energy ranges. The vertical extent of the source is given
as the width of the curve at the level of half the peak-flux.
It can also be seen how the position of the maximum is shifted
to lower heights for increasing energy.
Fig.~\ref{plotsix} illustrates the expected vertical extend in the thick-target
model compared to the observations for selected events.
In the following we are discussing the individual events in more detail.
\subsection{2002 May 31}
\begin{figure}[h]
\center{\resizebox{0.85\hsize}{!}{\includegraphics{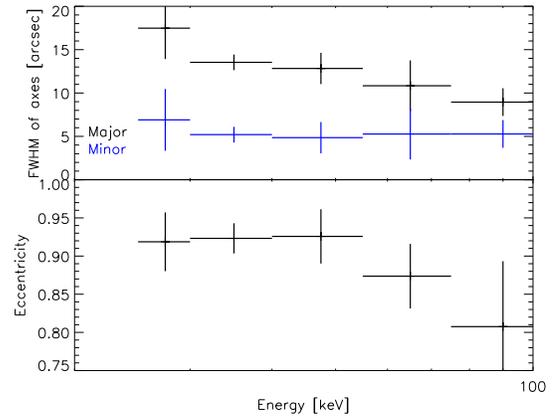}}}
\caption {Major and minor axes (top), eccentricity (bottom) in the event of 2002 May 31 which was fitted with an ellipse. }
\label{plotseven}
\end{figure}

This is a limb event with only one footpoint (Fig.~\ref{plotone}, top left).
The event was best fitted with a single elliptical Gaussian. A decrease of the radial
position with energy is observed, although the individual errors on the position
are rather large (Fig.~\ref{plotfour}). We fitted the density model described in Eq.~\ref{denseq} finding a density scale height of $h_0=204 \pm 37$ km.
The observations also suggest a decreasing FWHM with energy,
although with large uncertainties.
This is the only event that had a clearly elliptical footpoint shape.
The major and minor axes as well as the eccentricity 
are displayed in Fig. \ref{plotseven}.
The major axis is oriented along the limb. The morphology suggests that the major
axis of the ellipse is perpendicular to the magnetic field of the loop,
while the minor axis is parallel to the field. The extent of the major 
axis decreases with increasing energy, which can be interpreted as the signature of electron
transport in a converging magnetic field. The minor axis is constant as a function
of energy and much larger than the size expected from the thick-target model.

\subsection{2003 April 26}
This event has two very compact, close footpoint sources (named north and south in Fig.~\ref{plotfour}) 
that were best fitted with two circular Gaussian sources. In both sources, a tendency 
to a smaller radial distance with increasing energy can be observed. The fitted density scale 
heights are $h_0=132\pm 38$ km for the northern footpoint and $h_0=181 \pm 40$ km 
for the southern footpoint. The FWHM of the northern footpoint is constant 
as a function of energy. The size of the southern footpoint decreases as a function of 
energy except for the measurement at highest energies, which might be due to low count rate 
in the source. Interestingly, the size at 40 keV to 60 keV of the northern footpoint 
is consistent with the thick-target prediction, but not at the other energies. 
This indicates that geometrical effects such as projection or footpoint motion 
are negligible in this case since they would affect all energy ranges equally.
\subsection{2004 January 7}
This flare happened one day after the 2004 January 6 event analyzed by \citet{Ko08,Koet10} 
in the same active region. It has two footpoints although one is much fainter than the other. 
A model of two circular Gaussians is fitted to include the emission of the weak footpoint 
but only the results for the stronger footpoint are shown in Fig.~\ref{plotfour} (top right) 
due to the large uncertainties resulting from the small count rates in the weaker footpoint. 
The density scale height was found to be $h_0=230\pm 40$~km. The size is constant as a function 
of energy except in the lowest energy band in which the size is considerably larger. 
Although the spectrum suggests purely non-thermal emission in this energy band, 
it cannot be ruled out entirely that there is still some coronal emission which 
might affect the fitted size.

\subsection{2005 July 13}
This event has two footpoints (named north and south in Fig.~\ref{plotfour}) of equal 
intensity that are best fitted with two circular Gaussians. The radial position of the 
southern footpoint clearly decreases with energy. A density scale 
height of $h_0=235\pm 56$ km is found in this case. The height of the northern footpoint 
source also decreases as a function of energy, except in the lowest energy band. 
The size of both footpoints can be considered constant.
\subsection{2005 August 22}
The event has two footpoints with the northern footpoint being much fainter 
than the southern footpoint. Two circular Gaussians were fitted, but only the results 
for the southern footpoint are shown in Fig.~\ref{plotfour}. There is an indication 
for a decrease of position with energy followed by an increase, 
although the uncertainties are rather large. FWHM size is approximately constant.
\subsection{2005 August 23}
Only one distinct footpoint is measurable in this event although imaging over a larger 
energy band suggests the existence of a second, very faint footpoint. One circular Gaussian 
source was the best model for this case. Within the uncertainties, radial position 
as well as FWHM are constant.

\section{Physical interpretation of the observations}\label{physics}
While some of the described events display a clear decrease of radial distance 
with energy (similar to previous results) constant positions are also observed. 
In the classical thick-target model, one would expect a decrease of the height 
of a source with increasing energy. There are a number of effects that 
can influence the measured positions and sizes.
\subsection{Projection effects}\label{pro} 
The most obvious is projection effects. 
In observations we measure the position of the source in radial direction from the center 
of the Sun. For sources at or close to the limb, these positions are directly related 
to the height of the source above the photosphere. Closer to the center of the Sun, 
sources at different heights in the chromosphere will be seen in projection 
on top of each-other. This would result in an observed constant radial position 
as a function of energy. To avoid this we focused on near limb events. However, even for near limb events, an effect due to the heliocentric angle is expected. Table 1 lists the cosine of the heliocentric angles of the flaring region. The 2005 August 22 event is the furthest away from the limb at a heliocentric angle of $\theta=62\degr$, corresponding to $\mu=\cos(\theta)=0.47$. This introduces a factor $1/\sqrt{1-\mu^2}$ to the effective heights $h_{true}=h_{fitted}/\sqrt{1-\mu^2}$ which could affect the density fits. While no density fit was possible for the 2005 August 22 event ($\sqrt{1-\mu^2}=0.88$) because the positions as a function of energy were constant within uncertainty, in all the other events $\sqrt{1-\mu^2}\ge 0.96$ and the resulting effect is smaller than the uncertainties of the fits. Therefore, the influence of the heliocentric angle on the measured source heights and derived densities is negligible. The contribution of projection to the measured sizes can be estimated using a simple geometrical model. For a source with fitted horizontal extent (major axis) $a$ the fitted vertical extent (minor axis) $b$ is composed of:
\begin{equation}
b=b_{true}\times\sqrt{(1-\mu^2)}+a\times \mu
\end{equation} \label{abmu}
where $b_{true}$ is the true vertical extent. For events exactly at the limb $\mu=0$, thus $b=b_{true}$. Assuming a circular shape of the footpoint at a given height the extent perpendicular to the radial direction, which is unaffected by projection effects, can be used to estimate the true vertical size. Figure \ref{plotsix} illustrates the effect.

In all of the observed events, the measured source size is larger by at least a factor 
of three compared to the expected size in a thick-target model (Fig. \ref{plotsix}). Even when projection effects are included, the sizes are still more than a factor of 2 larger than expected from the thick target model.
Currently there is only one explanation i.e. multi-threaded loop density structure 
that has been investigated in more detail and that was used to explain the observed 
size in the event analyzed by \citet{Koet10}. Another possibility related to the density structure is a double-exponential density structure with a second, larger scale-height at higher altitudes as proposed by \citet{Sa10}. This might affect the height of the sources as a function of energy and possibly the sizes. However, it is unlikely that the effect will be large enough to explain the observed sizes. Moreover, in terms of the positions in the events presented here a density structure with a single scale-height is sufficient to explain the observed function of radial distance versus energy. However there are other effects 
such as magnetic mirroring, pitch angle scattering or X-ray albedo that
are expected to affect the size of the source.

\subsection{Magnetic mirroring}
In the standard thick-target model, magnetic mirroring is neglected. 
However, in a converging magnetic field it is expected that some particles 
are mirrored back from the footpoints, if they were injected at a pitch angle 
relative to the magnetic field lines. The mirroring point depends on the initial 
pitch angle of the electrons but is independent on the electron energy. 
The bulk of the emission is expected to originate from the densest part 
of the chromosphere to which the electrons are able to penetrate. 
A single mirroring point would lead to a constant position as a function 
of energy such as observed in the 2005 August 23 event. A behavior as in the stronger 
source of the 2003 April 26 event could also be envisaged. If the stopping depth 
for low energetic electrons is substantially higher than the mirroring point, 
those electrons will encounter a thick-target, while higher energetic electrons 
will be mirrored. Therefore, a decrease of radial position with energy will be 
observed at low energies, a constant position at higher energies. The size of 
HXR sources will depend on the pitch angle spread of the electrons. Field aligned 
electrons will penetrate deeper while electrons with a large pitch angle will 
be mirrored higher. This would make the source larger. On the other hand, 
in the extreme case of injection at a 90$\degr$ angle to the magnetic field 
the electrons will stay at the height of the injection point, 
gradually losing energy. This would result in a constant position 
and a source size determined by the size of the acceleration region.
\subsection{Collisional pitch angle scattering}
Another effect that might not be negligible is collisional pitch angle scattering. 
It is likely to increase the size of the source \citep{Co00}, but this increase is not expected to be large enough to explain the observations. Additional scattering due to various plasma waves can be anticipated \citep[e.g.][]{Bi10,Ha09, St02}. While this can be substantial, it is difficult to quantify the level of turbulence in a flaring 
atmosphere.

\subsection{X-ray albedo}
Photon backscattering (albedo) from the photosphere can change the observed position and size 
as shown by \citet{Ko10} and \citet{Ba11}. The additional albedo flux 
results in a shift of the position radially toward the disk center and in an increase 
of the observed source size by several arcseconds. Since albedo is energy dependent 
(with the strongest contribution between around 30 and 50 keV), the effect on the position and the size will be most pronounced at those intermediate energies and smaller at low and high energies. 
\citet{Ko10} further find that this effect can be important even for large heliocentric angles, 
i.e. events near the limb. Being the closest to the solar disk (at a cosine of the heliocentric 
angle of $\mu=0.47$), the observations of the event of 2005 August 22 might be affected by albedo. 
The positions would be shifted radially toward the disk center, most pronounced 
at energies between 30 and 50 keV. This could explain the observed pattern (Fig.~\ref{plotfour}, bottom left).

\section{Discussion and Conclusions} \label{discussion}
In this work we performed the first {multi-event study of both source positions and sizes 
of flare footpoints as a function of energy.
The radial positions as a function of energy in 4 out of 6 events follows a decreasing trend with 
increasing energy as expected in the classical thick-target model. The positions were fitted with 
an exponential density model, finding scale heights between 132 and 235 km, consistent with previous studies. 
The other two events show a weak dependence of the positions on energy. The vertical sizes range from 1.3 up to 8 arcsec, therefore being up to a factor 3 larger 
than expected from the simple thick-target model, even for the events with clear dependence 
of the position on the energy and after correcting for projection effects due to the heliocentric angle (see Fig.~\ref{plotsix}). While it has been argued, based 
on traditional imaging techniques, that RHESSI might simply under-resolve footpoint sizes \cite[e.g.][]{De09}, 
other studies suggest that X-ray footpoint sizes of the order of several arcseconds are real \citep{Ba11,Koet10}. This is supported by negligible modulation in the finest RHESSI grid (grid one), which gives
a lower limit of the source size of 3.9 arcsec in the events presented here.
Currently, the only explanation for this in the context of a simple thick-target model is a multi-threaded loop density structure. Here we discuss other mechanisms which might explain this intriguing finding. In the thick-target model, electrons are injected parallel to the magnetic field lines. In reality, injection at a pitch angle relative to the field lines can be expected.
In a converging magnetic field, the initial pitch angle distribution will be modified and magnetic mirroring will play a non-negligible role. Depending on the height of the mirroring point, the observed height as a function of energy will be constant. This can explain the observations of the position in two events. At the same time, the size is expected to be affected, as well. In addition collisional pitch angle scattering, as well as wave-particle interactions will further modify the pitch angle and therefore the observed size, although collisional pitch angle scattering alone will not be sufficient to enhance the size.
A third possibility is the influence of X-ray albedo. Although the effect is expected to be small in limb events, it might affect the measurements in the event of 22 August, which is closest to the disk center.

While the presented events are very individual, a general trend can still be found when averaging the sizes. Figure \ref{ploteight} shows the average vertical and horizontal size as a function 
of energy. In the case of a circular Gaussian source, the vertical and horizontal sizes are identical and equal to the FWHM, while in the case of the 2002 May 31 event, the minor and major axis were used. The figure suggests a decrease of both dimensions with energy.
\begin{figure}[h]
\center{\resizebox{0.85\hsize}{!}{\includegraphics{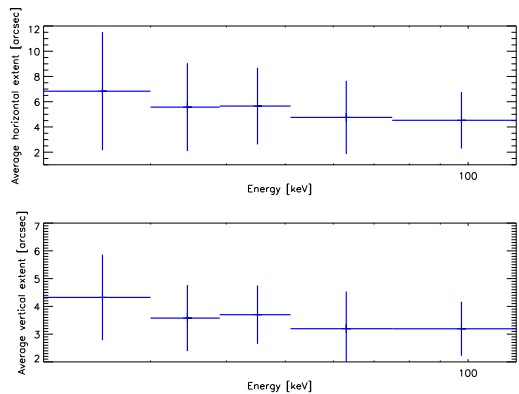}}}
\caption {Average horizontal (top) and vertical (bottom, projection corrected) extent of the sources. The error bars indicate the standard deviation of the scatter of the individual points. }
\label{ploteight}
\end{figure}
In order to understand the physics and explain the observations, one needs to carefully address both the position and the sizes since most of the effects discussed are expected to affect both, size and position. Test particle simulations including magnetic field, collisional pitch angle scattering as well as other effects such as pitch angle scattering due to turbulence are necessary and have to be compared with the observations. Such a complete treatment will make it possible to use X-ray footpoints as an independent means of determining the chromospheric magnetic field and density structure.
\acknowledgments
The authors wish to thank Gordon Hurford
and Richard Schwartz for helpful discussions.
This work is supported by the Leverhulme Trust (M.B., E.P.K.),
STFC rolling grant (E.P.K.) and STFC Advanced Fellowship (E.P.K.).
Financial support by the European Commission
through the SOLAIRE (MTRN-CT-2006-035484)
and the HESPE networks is gratefully acknowledged.
\bibliographystyle{apj}
\bibliography{mybib}

\end{document}